\documentclass[aip,jap,numerical,floatfix,preeprint]{revtex4-1}
\usepackage[utf8]{inputenc}  
\usepackage[T1]{fontenc}  
\usepackage[english]{babel} 
\usepackage{epsfig} 
\usepackage{parskip}
\usepackage{setspace}
\usepackage{caption3} 
\usepackage{subfig}
\usepackage{ifpdf}
\usepackage{nomencl}
\usepackage{empheq}
\usepackage{natbib}
\DeclareMathSizes{10}{10}{10}{10}
\addto\captionsenglish{\renewcommand{\figurename}{FIG.}}
\begin{document}
\title{A simple radiative thermal diode}
\author{Elyes NEFZAOUI$^{*}$, Jérémie DREVILLON, Younès EZZAHRI and Karl JOULAIN$^{**}$}
\address{$^{1}$Institut Pprime, CNRS-Universit\'e de Poitiers-ENSMA, D\'epartement Fluides, Thermique, Combustion, ENSIP-B\^atiment de m\'ecanique, 2, Rue Pierre Brousse, F 86022 Poitiers, Cedex, France \\
\small Corresponding authors : $^{*}$elyes.nefzaoui@univ-poitiers.fr, $^{**}$karl.joulain@univ-poitiers.fr}
\begin{abstract}
We present a thermal rectification device concept based on far-field radiative exchange between two selective emitters. Rectification is achieved due to the fact that one of the selective emitters radiative properties are independent on temperature whereas the other emitter properties are strongly temperature dependent. A simple device constituted by two multilayer samples made of metallic (Au) and semiconductor (Si and HDSi) thin films is proposed. This device shows a rectification up to $70\%$ with a temperature difference $\Delta T = 200$ K, a rectification ratio that has never been achieved so far with radiation-based rectifiers. Further optimization would allow larger rectification values. Presented results might be useful for energy conversion devices, smart radiative coolers / insulators engineering and thermal modulators development.
\end{abstract}
\maketitle
Thermal rectification can be defined as an asymmetry in the heat flux when the temperature difference between two interacting thermal reservoirs is reversed. The realization of a device exhibiting such an uncommon behavior, a thermal diode for instance, would pave the way to the development of thermal circuits in the manner non-linear electronic devices marked the genesis of modern electronics\cite{Wang2008}. Consequently, an increasing interest has been given to thermal rectifiers during the last years.
Experimental rectifiers based on carbon and boron nitride nanotubes \cite{Chang2006}, semiconductor quantum dots \cite{Scheibner2008} and bulk cobalt oxides\cite{Kobayashi2009} have been realized. Besides, numerous theoretical models have also been proposed, based on non-linear lattices \cite{Terraneo2002,Li2004,Li2005,Hu2006}, graphene nano-ribbons\cite{Hu2009,Yang2009} and several other interesting mechanisms \cite{Segal2008,Yang2007}. Some authors went beyond the thermal rectifiers issue and proposed theoretical models of thermal logical gates \cite{Wang2007} and a thermal transistor \cite{Lo2008}.
An overwhelming majority of the proposed devices schemes are based on conductive heat transfer channels control and very few radiative thermal rectifiers have been reported thus far. 
Indeed, a theoretical study and an experimental suggestion of a radiative thermal rectifier based on non-linear solid-state quantum circuits operating at very low temperatures (a few mK) have recently been presented\cite{Ruokola2009}. A rectification up to $10\%$ is predicted. Moreover, two theoretical schemes of radiative thermal rectification based on near-field thermal radiation control have lately been proposed \cite{Otey2010,basu2011}. A rectification up to $40\%$ is theoretically predicted. 

In this letter, a far-field radiation based thermal rectifier made of selective emitters previously developed for thermo-photovoltaic applications\cite{Nefzaoui2012} is presented. This device provides larger rectification ratios than reported in literature with radiative thermal rectifiers and comparable temperature differences. Besides, the far-field radiation based devices are more suitable for experimental implementations. Finally, since the presented device is based on spectrally tunable selective emitters, the proposed rectifier allows high rectification ratios in different spectral ranges, thus at different operating temperatures.

\begin{figure}[ht]
\begin{center}
\includegraphics[width=0.3\textwidth]{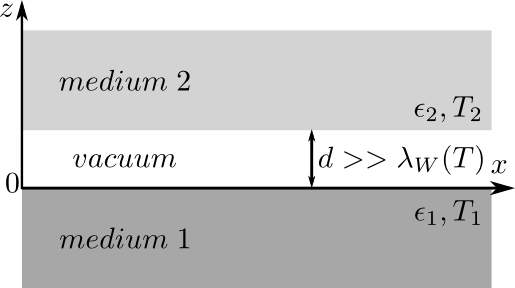}
\caption[]{Two parallel planar bodies separated by a distance $d$ much larger than Wien wavelength $\lambda_W(T)$. In forward bias configuration, $T_1 = T_h =  500$ K and $T_2 = T_c = 300$ K. In reverse bias configuration, $T_1 = T_c$ and $T_2 = T_h$.}
\label{fig:PlanePlane}
\end{center}
\end{figure}

\begin{figure}[ht]
\begin{center}
\includegraphics[width=0.7\textwidth]{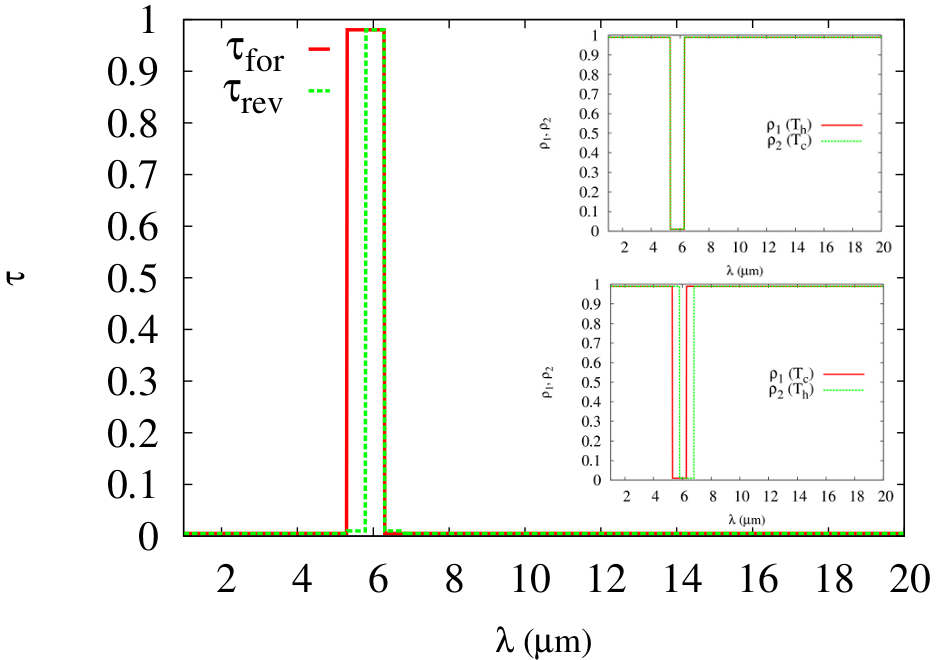}
\caption{The transmission coefficient in forward and reverse biased configurations and bodies $1$ and $2$ reflectivities in forward (top inset) and reverse bias (bottom inset) configurations with $\Delta \lambda = 1$ $\mu$m and $\rho_{min} = 10^{-2}$. In both configurations, $\lambda_{p,1} (T) = \lambda_W (T_h = 500K)= 5.8 $ $\mu$m while $\lambda_{p,2} (T_c) = \lambda_{p,1}$ in forward bias and $\lambda_{p,2} (T_h) = \lambda_{p,1} +  \frac{\Delta \lambda}{2}$ in reverse bias.}
\label{fig:RTau_th}
\end{center}
\end{figure}

Figure \ref{fig:PlanePlane} presents a schematic of the proposed device composed of two parallel planar bodies $1$ and $2$ separated by a gap of thickness $d$ and characterized by their dielectric functions and temperatures ($\varepsilon_1(T_1),T_1$) and ($\varepsilon_2(T_2),T_2$) respectively.
We note $\dot{q}_{for}$ and $\dot{q}_{rev}$ the net exchanged heat flux densities between $1$ and $2$ in forward ($T_1 = T_h =  500$ K and $T_2 = T_c = 300$ K) and reverse bias ($T_1 = T_c$ and $T_2 = T_h$) configurations, respectively. The rectification coefficient $R$ defined as the relative difference between these two quantities is then given by:
\begin{eqnarray}
R & = & \frac{\dot{q}_{for} - \dot{q}_{rev}}{max(\dot{q}_{for},\dot{q}_{rev})}
\end{eqnarray}
The two bodies are assumed to be in vacuum. $\dot{q}$ is thus a radiative heat flux (RHF). The considered bodies radiative properties, in particular their emissivities and reflectivities ($\epsilon$ and $\rho$ respectively), are completely governed by 
their dielectric functions and geometries. In the case of opaque bodies, energy conservation and Kirchhoff's laws combination leads to the following relation between the monochromatic emissivity and reflectivity at a given temperature:
\begin{eqnarray}
\epsilon (T,\lambda) & = & 1 - \rho(T,\lambda)
\label{eq:EnergyConservation}
\end{eqnarray}
We also assume the two bodies are lambertian sources. $\epsilon$ and $\rho$ are thus direction-independent. \\
The gap width $d$ is assumed to be much larger than the dominant thermal radiation wavelength (Wien wavelength) $\lambda_W(T) = \frac{hc}{k_bT}$ where $h$, $c$, $k_b$ and $T$ are Planck constant, the speed of light in vacuum, Boltzmann constant and the absolute temperature, respectively. The net RHF density exchanged by the two media resumes then to the far field contribution which can be written \cite{Modest1993}:
\begin{eqnarray}
\dot{q} (T_1,T_2) & = & \pi \int_{\lambda=0}^{\infty} [I^0(\lambda,T_1)-I^0(\lambda,T_2)] 
\tau(\lambda,T_1,T_2) d \lambda
\label{eq:FluxDensity}
\end{eqnarray}
where 
\begin{eqnarray}
I^0(\lambda,T) & = & \frac{h c^2}{\lambda^5}\frac{1}{e^{\frac{h c}{\lambda k_b T}}-1}
\label{eq:Chap5-PlanckBBIntensity}
\end{eqnarray}
is the black body intensity at a temperature $T$ and
\begin{eqnarray}
\tau(\lambda,T_1,T_2) & = & \frac{\epsilon_1 (\lambda, T_1) \epsilon_2 (\lambda, T_2)}{1-\rho_1 (\lambda, T_1) \rho_2 (\lambda, T_2)}
\label{eq:TransmissionCoeff}
\end{eqnarray}
is the monochromatic RHF density transmission coefficient between $1$ and $2$.
Now, let us assume the two bodies are selective emitters so that they behave as quasi-monochromatic spectral emitters, i.e. they present reflectivity valleys (or emissivity peaks) of the same finite width $\Delta \lambda$ at given wavelengths, $\lambda_{p,i}, i \in \{1,2 \}$ for instance.
Their reflectivities can thus be defined as follows:
\begin{eqnarray}
\rho_i (\lambda,T) = \left\{
    \begin{array}{ll}
        \rho_{min} \simeq 0 & \mbox{if } \lambda \in [\lambda_{p,i} (T)-\frac{\Delta \lambda}{2}, \lambda_{p,i} (T)+ \frac{\Delta \lambda}{2}] \\
        \rho_{iI} \simeq 1 & \mbox{if } \lambda < \lambda_{p,i} (T)-\frac{\Delta \lambda}{2} \\
        \rho_{iII} \simeq 1 & \mbox{if } \lambda > \lambda_{p,i} (T)+\frac{\Delta \lambda}{2}
    \end{array}
\right.
\label{eq:Epsilon}
\end{eqnarray}
These reflectivities are illustrated in Fig. \ref{fig:RTau_th} insets for a given set of parameters : $\Delta \lambda = 1$ $\mu$m, $\rho_{min} = 10^{-2}$, $\rho_{iI} = \rho_{iII} =  1 - \rho_{min}$ and $\lambda_{p,1} (T) = \lambda_W (T_h = 500K)$.
Body $1$ emissivity is T-independent. Body $2$ emissivity is initially identical to that of body $1$ in forward bias ($\lambda_{p,2} (T_c) = \lambda_{p,1}$) while its peak shifts to the red by $\delta_{rev} \lambda_p$ when the temperatures of the two bodies are reversed i.e. $\lambda_{p,2} (T_h) = \lambda_{p,1} + \delta_{rev} \lambda_p$. This shift with the temperature of one of the selective emitters radiative properties is the key to achieve thermal rectification.
Consider for now a small shift verifying $\delta_{rev} \lambda_p < \Delta \lambda$ thus the two peaks still partially overlap in reverse biased configuration. In Fig. \ref{fig:RTau_th}, $\delta_{rev} \lambda_p = \frac{\Delta \lambda}{2}$. 
The transmission coefficient resulting from this emissivities choice for both configurations is plotted in Fig. \ref{fig:RTau_th}. We can then note a one-$\mu$m wide spectral transmission window of RHF in forward bias and a narrower window in reverse bias. This asymmetry when the temperatures of the two bodies are inverted would lead to a thermal rectification. 
In fact, the exchanged RHF in forward bias is given by :
\begin{eqnarray}
\dot{q}_{for} & \simeq & \sigma T_h^4 \left[ F_{\lambda_{p,1}+\frac{\Delta \lambda}{2}}(T_h) - F_{\lambda_{p,1}-\frac{\Delta \lambda}{2}} (T_h) \right] 
- \sigma T_c^4 \left[ F_{\lambda_{p,1}+\frac{\Delta \lambda}{2}}(T_c) - F_{\lambda_{p,1}-\frac{\Delta \lambda}{2}} (T_c) \right] \nonumber \\
& + & \pi \int_{\lambda \in \mathcal{L}_{for}} \frac{\epsilon_{1} (\lambda,T_h) \epsilon_{2} (\lambda,T_c)}{1-\rho_{1} (\lambda,T_h) \rho_{2} (\lambda,T_c)} \left[ I^0(\lambda,T_h) - I^0(\lambda,T_c) \right]
\label{eq:PiieceWiseForRHF}
\end{eqnarray}
where 
\begin{eqnarray}
\sigma T^4 & = & \pi \int_0^{\infty} I^0(\lambda^{\prime},T) d\lambda^{\prime}
\end{eqnarray}
$\sigma = \frac{k_b^4 \pi^2}{60 \hbar^3 c^2}$ is Stefan-Boltzmann constant,
\begin{eqnarray}
F_{\lambda} (T) & = & \pi \frac{\int_0^{\lambda} I^0(\lambda^{\prime},T) d\lambda^{\prime}}{\sigma T^4}
\end{eqnarray}
is the spectral fraction of black body radiation emitted in the spectral range $\lambda^{\prime} \leq \lambda$ and $\mathcal{L}_{for} = [0,\lambda_{p,1}-\frac{\Delta \lambda}{2}] \cup [\lambda_{p,1}+\frac{\Delta \lambda}{2},\infty]$.
On the other hand, the exchanged RHF in reverse bias is given by :
\begin{eqnarray}
\dot{q}_{rev} & \simeq & \sigma T_h^4 \left[ F_{\lambda_{p,1}+\frac{\Delta \lambda}{2}}(T_h) - F_{\lambda_{p,1}+\delta_{rev} \lambda_p-\frac{\Delta \lambda}{2}} (T_h) \right]
- \sigma T_c^4 \left[ F_{\lambda_{p,1}+\frac{\Delta \lambda}{2}}(T_c) - F_{\lambda_{p,1} + \delta_{rev} \lambda_p-\frac{\Delta \lambda}{2}} (T_c) \right] \nonumber \\
& + & \pi \int_{\lambda \in \mathcal{L}_{rev}} \frac{\epsilon_{1} (\lambda,T_h) \epsilon_{2} (\lambda,T_c)}{1-\rho_{1} (\lambda,T_h) \rho_{2} (\lambda,T_c)} \left[ I^0(\lambda,T_h) - I^0(\lambda,T_c) \right]
\label{eq:PiieceWiseRevRHF}
\end{eqnarray}
where $\mathcal{L}_{rev} = [0,\lambda_{p,1}+\delta_{rev} \lambda_p-\frac{\Delta \lambda}{2}] \cup [\lambda_{p,1}+\frac{\Delta \lambda}{2},\infty]$.

The piecewise decompositions of the exchanged RHF in Eqs. \ref{eq:PiieceWiseForRHF} and \ref{eq:PiieceWiseRevRHF} take into account the fact that $\epsilon_i \simeq 1$ within the structure $i$ peak. Since $\epsilon_{i} \simeq 0$ elsewhere else, the terms with the pre-factor $\frac{\epsilon_1 \epsilon_2}{1 - \rho_1 \rho_2}$ in the integrand are of the first order. In reverse biased configuration, we can note that the spectral domains of the order zero and the first order terms are reduced and extended respectively due to the shift between the two peaks. The reduction of the dominating term (order $0$) due to its spectral domain diminution, if it is different from the higher order term ($1$st order) increase, would lead to an asymmetry of the RHF and therefore a rectification phenomenon.\\
For the above presented situation, calculations lead to: $\dot{q}_{for} = 122.67 $ W.m$^{-2}$, $\dot{q}_{rev} = 63.97 $ W.m$^{-2}$ and $R =  0.48$.
This rectification value is of the order of the largest reported values for radiative thermal rectifiers. According to the previous assumptions, this rectification ratio obviously depends on the chosen parameters, in particular the temperatures $T_h$ and $T_c$, the peaks position $\lambda_p$ compared to $\lambda_W (T_h)$, their width $\Delta \lambda$ compared to the black body spectrum useful width at the largest temperature, the shift between the two peaks in reverse bias $\delta_{rev} \lambda_p$ compared to $\Delta \lambda$ and the reflectors and emitters quality characterized by $\rho_{ij}$ and $\rho_{min}$, respectively. 
It is clear that the maximal rectification ratio is reached when the peaks shift in reverse bias is larger than the peaks width and when the bodies are perfect reflectors outside the peaks, i.e. when $ \delta_{rev} \lambda_p / \Delta \lambda \geq 1$ and $\rho_{ij} = 1$ respectively : reverse bias RHF vanishes and rectification reaches 1.
The first condition on $\delta_{rev} \lambda_p$ is more easily realized with narrow emissivity peaks. However, narrower emissivity peaks would decrease forward bias RHF density which makes the rectification ratio more sensitive to the RHF density noise (high order terms), i.e. to heat transmitted in a spectral range outside the peaks thus to $\rho_{ij}$.

Now, we propose a real system that would allow a practical realization of the above presented device concept. 
The key point to obtain such a rectification device is to use selective emitters i.e bodies that behave as mirrors in the mid-infrared except on a narrow-band wavelength range where they strongly emit. One of the bodies has to have temperature independent radiative properties whereas the other one must present an emission peak shift sufficiently large with temperature. 
For this purpose, we use selective emitters based on Fabry-Pérot resonant cavities previously developed for thermo-photovoltaic applications \cite{Nefzaoui2012}. First, these structures present emissivity peaks in the near and mid infra-red. Second, the peaks positions and widths can be controlled by few simple parameters such as layers thicknesses and the cavity reflectors quality. Finally, the use of selective emitters with different materials which dielectric functions have sufficiently different T-dependencies might allow the observation of a rectification phenomenon.

\begin{figure}[ht]
\begin{center}
\includegraphics[width=0.35\textwidth,height=0.25\textheight]{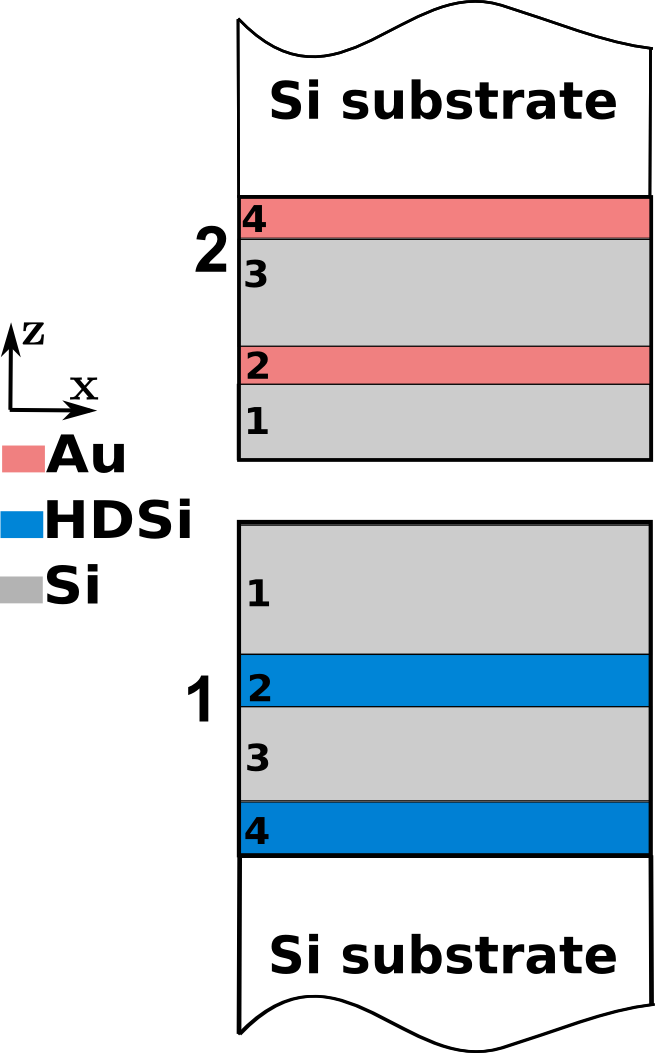}
\caption{
A radiative thermal rectification device based on two Fabry-Pérot cavities. The reflectors of body 1 are made of \textit{p}-type heavily doped silicon (HDSi) at $N=10^{20}$ cm$^{-3}$ while those of body 2 are made of gold (Au). The transparent layers and the substrates are made of intrinsic silicon. The dimensions of the different layers in nanometer are $d_{1,1}=350$, $d_{1,2}=100$, $d_{1,3}=190$ and $d_{1,4}=100$ for body $1$, and  $d_{2,1}=90$, $d_{2,2}=40$, $d_{2,3}=250$, $d_{2,4}=40$ for body $2$. In forward bias $T_1 = T_h$ and $T_2 = T_c$. In reverse bias, the two bodies temperatures are inverted.}
\label{fig:device}
\end{center}
\end{figure}

\begin{figure}[ht]
\begin{center}
\includegraphics[width=0.7\textwidth]{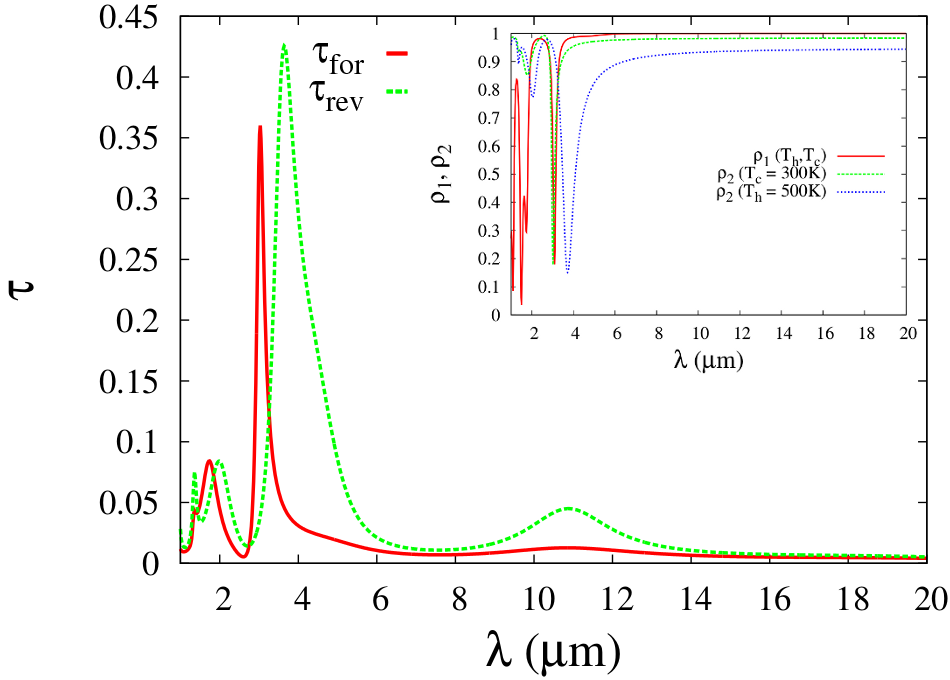}
\caption[]{Spectral transmission coefficient of the device presented in Fig. \ref{fig:device} in forward ($\tau_{for}$) and reverse bias ($\tau_{rev}$) configurations. The inset shows the reflectivity of the two bodies in both configurations. In red, the reflectivity of body $1$ at $T_h$ and $T_c$ which is $T$-independent. For body $2$, in green and blue the reflectivity at $T_c$ (forward) and $T_h$ (reverse) respectively.}
\label{fig:TAUdevice}
\end{center}
\end{figure}

Consider the system presented in Fig. \ref{fig:device}.
The interacting bodies are two Fabry-Pérot cavities made of four thin films alternating a transparent and a reflecting film on a thick intrinsic silicon substrate. Transparent layers are made of intrinsic silicon. Reflectors are made of \textit{p}-type heavily doped silicon (HDSi) in one body, and gold (Au) in the other.
Gold and heavily doped silicon dielectric functions at considered temperatures are modeled by Drude models (See appendix). We assume the system geometry unchanged in the considered temperature range. In fact, the linear thermal expansion coefficient of used materials is, at the most, of the order of $10^{-5}$ K$^{-1}$. This leads, for a $100$ K temperature variation, to a relative variation of films thicknesses of one thousandth. Such a variation has no significant impact on the considered structures radiative properties\cite{Nefzaoui2013}. 
These two structures reflectivities at normal incidence at $T_c=300$ K and $T_h = 500$ K calculated by the transfer matrix method are presented in the inset of Fig. \ref{fig:TAUdevice}. Only normal reflectivity is plotted since these structures show isotropic radiative properties\cite{Nefzaoui2012}. We can note that Si/Au structure reflectivity varies rapidly with temperature while Si/HD-Si structure reflectivity shows no significant variation.
The impact of this T-dependence on the exchanged RHF can be deduced from the plot of the spectral transmission coefficient $\tau$ in Fig. \ref{fig:TAUdevice}. In reverse bias configuration, $\tau$ peak magnitude is increased, the peak shifts to the red, i.e. closer to the black body maximal emission wavelength at $T_h = 500$ K and gets wider. In addition, a slightly wider and smaller secondary peak appears at larger wavelengths. 
All these evolutions will obviously induce an increase in the net RHF. 
The numerical application leads to $\dot{q}_{for} = 15.5 $ W.m$^{-2}$, $\dot{q}_{rev} = 52.3 $ W.m$^{-2}$ and $R =  - 0.7$. \\
The proposed device rectification is therefore of $70\%$, a value larger by almost a factor of two than the largest reported values for radiative rectifiers\cite{Otey2010}. However, instead of diminishing, the net RHF density increases when switching from forward to reverse bias configurations. This is mainly due to the fact that, in addition to its emissivity peak shift at high temperature, the gold-composed Fabry-Pérot cavity becomes less reflecting outside the peak which allows it to exchange more heat over the whole black body spectrum. The RHF density drop due to the peaks mismatch is compensated by the higher emissivity / absorptivity of the gold-silicon selective emitter over the considered spectral range at $T_h = 500$ K. According to Eq. \ref{eq:PiieceWiseRevRHF}, this effect is mainly due to the fact that high order terms become dominating when the temperatures of the two bodies are reversed. Note that the proposed device can be improved to obtain higher rectification. Further optimization of the presented system with a strong constraint on the reflectors quality, i.e. on the values of $\rho_{ij}$, would lead to performances comparable to those of the ideal device. 

In this letter, we presented a theoretical concept of a thermal rectifier based on far field thermal radiation. Then, we suggested a real nano-structured system based on two multi-layered bodies as a candidate for an experimental observation. This system shows an absolute rectification value larger by almost a factor two than those proposed in literature with radiation-based rectification devices and comparable temperature differences\cite{Otey2010}. Note that the proposed device is quite easy to realize since it involves samples radiating in the far-field.   
Finally, an optimization process would certainly lead to larger rectification values. The use of very high thermal expansion coefficient meta-materials would offer new possibilities to control the peaks mismatch, thus the rectification ratio, through the cavity width variation.
Presented results might be useful for energy conversion devices, smart radiative coolers / insulators engineering and thermal modulators development.
\section*{Acknowledgments}
Authors gratefully acknowledge the support of the Agence Nationale de la Recherche through the Source-TPV Project No. ANR 2010 BLAN 0928 01. 
\section*{Appendix : Materials optical properties temperature dependence}
Materials optical properties temperature dependence is a key point in the radiative thermal diode device we propose. This appendix details the temperature dependence modeling we used in our calculations for gold, heavily doped silicon and intrinsic silicon optical properties.
\subsection{Gold}
Gold (Au) dielectric function can be modeled by a Drude model\cite{Pells1969}.
\begin{eqnarray}
\epsilon(\omega) & = & 1 - \frac{\omega_p^2}{\omega(\omega+\imath \Gamma)}
\label{eq:drude}
\end{eqnarray} 
where $\omega=2 \pi c / \lambda$ is the circular frequency, $\omega_p$ is the plasma frequency and $\Gamma$ is the damping coefficient.
The values of $\omega_p$ and $\Gamma$ for different temperatures are given in table \ref{tab:GoldDrudeTParam}\cite{Pells1969} for the spectral range $1-30$ $\mu$m.
\begin{table}[h!]
\begin{center}
\begin{tabular}{ccccc}
\hline
$T$ & $\sigma_0$ & $\tau$ & $\omega_p$ & $\Gamma$ \\
\footnotesize (K) & \footnotesize ($10^{17}$ s$^{-1}$) & \footnotesize ($10^{14}$ s) & \footnotesize ($10^{16}$ s$^{-1}$) & \footnotesize ($10^{13}$ s$^{-1}$) \\
\hline 295 & 3.79 & 2.4 &  1.4087 & 4.1667 \\
300$^*$ & 3.75 & 2.4 & 1.4099 & 4.2517 \\
470 & 2.35 & 1.4 & 1.4524 & 7.1429 \\ 
500$^*$ & 2.23 & 1.3 & 1.4502 & 7.634 \\
670 & 1.58 & 0.96 & 1.4381 & 10.417 \\ 
\hline
\end{tabular}
\caption{Drude model parameters for gold at different temperatures. The values of $\sigma_0$ (steady-state electrical conductivity) and $\tau$ (free electron relaxation time) are retrieved from literature\cite{Pells1969} while those of $\omega_p$ and $\Gamma$ are obtained through expressions \ref{eq:DrudeSigmaOmegaP} and \ref{eq:DrudeGammaTau}. \\ \footnotesize $^*$ : values are obtained by a linear interpolation (not from Ref\cite{Pells1969}.)}
\label{tab:GoldDrudeTParam}
\end{center}
\end{table}
The different parameters of table \ref{tab:GoldDrudeTParam} are related by \cite{Ashcroft1976}:
\begin{eqnarray}
\omega_p^2 = 4 \pi \frac{\sigma_0}{\tau}
\label{eq:DrudeSigmaOmegaP}
\end{eqnarray} 
and
\begin{eqnarray}
\Gamma = \frac{1}{\tau}
\label{eq:DrudeGammaTau}
\end{eqnarray} 
\subsection{Heavily doped silicon}
Heavily doped silicon dielectric function can also be model by a Drude model \cite{Marquier2004b, Basu2010a}:
\begin{eqnarray}
\epsilon(\omega) & = & \epsilon_{\infty} - \frac{\omega_p^{2}}{\omega(\omega+\imath \Gamma)}
\label{eq:drude}
\end{eqnarray} 
where $\epsilon_{\infty} = 11.7$ is the high frequency limit of the dielectric function. $\omega_p$ and $\Gamma$ are given by :
\begin{eqnarray}
\omega_p^{2} & = & \frac{N e^2}{(m^{*} \epsilon_0)}
\label{eq:chap4-FreqPlasma}
\end{eqnarray}
and
\begin{eqnarray}
\Gamma & = & \frac{1}{\tau} = \frac{e}{m^{*} \mu}
\label{eq:chap4-DrudeDamping}
\end{eqnarray} 
where $\epsilon_0$ is the vacuum permittivity. $\tau$, $\mu$ et $m^{*}$ are the relaxation time, mobility and effective mass of carriers (electrons / holes) respectively which depend on the carriers density $N$ and temperature \cite{Basu2010a, Marquier2004b, Green1990, Arora1982}. At the considered doping level, we assume complete impurity ionization.
The mobility of electrons and holes of phosphorus doped silicon as a function of carriers concentration and temperature is given by\cite{Arora1982} :
\begin{eqnarray}
\mu_e & = & 88 \; T_n^{-0,57} + \frac{7,4 \times 10^8  \; T^{-2,33}}{1 + [N / (1,26 \times 10^{17} \; T_n^{2,4})]0,88  \;  T_n^{-0,146}} \nonumber \\
\label{eq:ElectronMobilityTNDependance}
\end{eqnarray}
for electrons and
\begin{eqnarray}
\mu_h & = & 54,3 \; T_n^{-0,57} + \frac{1,36 \times 10^8  \;  T^{-2,23}}{1 + [N / (2,35 \times 10^{17}  \; T_n^{2,4})]0,88  \;  T_n^{-0,146}} \nonumber \\
\label{eq:HoleMobilityTNDependance}
\end{eqnarray}
for holes where $T_n = T / 300$ denotes the reduced temperature. These expressions show a very good agreement with experimental values up to $N = 10^{20}$ cm$^{-3}$.\\
The effective mass of electrons at room temperature is given by $m^* = 0.27m_0$ and $m^* = 0.34m_0$ for \textit{n}-type and \textit{p}-type doped silicon respectively, where $m_0 = 9.1 \times 10^{-31}$ Kg denotes the free electron mass in vacuum \cite{Marquier2004b}. The effective mass varies very slightly with temperature \cite{Green1990,Fu2006,Spitzer1957}. It is therefore considered temperature independent in this study. According to the given model and assumptions, the damping coefficient increases by almost $30\%$ (Table \ref{tab:DopedSiMobility}) while the plasma frequency remains constant in the considered temperature range.
\begin{table}[h!]
\begin{center}
\begin{tabular}{ccc}
\hline 
$T$ (K) & $300$ & $500$ \\ 
\hline $\mu$ (cm$^2$/V.s) & $55.38$  & $41.84$ \\ 
$\Gamma$ ($10^{12}$ s$^{-1})$ & $5.26$  &  $6.96$ \\ 
\hline
\end{tabular}
\end{center} 
\caption{Carriers mobility and damping coefficient for \textit{p}-type doped silicon at $N = 10^{20}$ (cm$^{-3}$) at $T = 300$ K and $T = 500$ K.}
\label{tab:DopedSiMobility}
\end{table}
\subsection{Intrinsic silicon}
Intrinsic silicon refractive index $n$ is assumed to be constant in the considered temperature range\cite{Ravindra1998,Sato1967} ($n = 3.42$). As a matter of fact, its value for slightly doped silicon at different temperatures is presented in table \ref{tab:SiRefractiveIndex-T} and shows very small variations. 
\begin{table}[ht]
\begin{center}
\begin{tabular}{ccc}
Doping type & $T$ (K) & $n$\\ 
\hline p & 331  & 3.3\\ 
 p & 547  & 3.4\\ 
\hline n & 293  & 3.43\\ 
 n & 543  & 3.48\\ 
\hline
\end{tabular}
\end{center} 
\caption{Mean values of $p$ and $n$-type slightly doped silicon refractive index $n$ in the spectral range $[1,6]$ $\mu$m at $T = 300$ K and $T = 500$ K.}
\label{tab:SiRefractiveIndex-T}
\end{table}

\end{document}